\documentclass{article}

\usepackage{arxiv}

\usepackage[utf8]{inputenc} 
\usepackage[T1]{fontenc}    
\usepackage{url}            
\usepackage{booktabs}       
\usepackage{amsfonts}       
\usepackage{nicefrac}       
\usepackage{microtype}      
\usepackage[colorlinks = TRUE, linkcolor=blue,citecolor=blue,urlcolor=blue]{hyperref}
\usepackage{doi}

\usepackage{mathrsfs, amssymb, amsmath, amsthm, mathtools, graphicx, subfigure, psfrag, rotating, pdflscape, url, colortbl, multirow}
\usepackage[authoryear]{natbib}
\usepackage{cleveref}       

\title{Parametric Modal Regression with Error in Covariates}

\date{\today}

\author{ \href{https://orcid.org/0000-0003-3265-6330}{\includegraphics[scale=0.06]{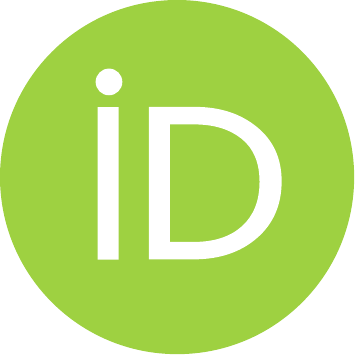}\hspace{1mm}Qingyang Liu} \\
	Department of Statistics\\
	University of South Carolina\\
	Columbia, SC 29201 \\
	\texttt{qingyang@email.sc.edu} \\
	\And
	\href{https://orcid.org/0000-0001-7077-0869}{\includegraphics[scale=0.06]{orcid.pdf}\hspace{1mm}Xianzheng Huang} \\
	Department of Statistics\\
	University of South Carolina\\
	Columbia, SC 29201 \\
	\texttt{huang@stat.sc.edu} \\
}

\renewcommand{\shorttitle}{Parametric Modal Regression with Error in Covariates}

\hypersetup{
	pdftitle={Parametric Modal Regression with Error in Covariates},
	pdfsubject={statistics},
	pdfauthor={Qingyang Liu and Xianzheng Huang},
	pdfkeywords={Beta distribution, bootstrap, corrected score, $M$-estimation, model misspecification},
}

\usepackage{fancyvrb}
\theoremstyle{plain}

\theoremstyle{definition}
\usepackage[]{graphicx}
\chardef\bslash=`\\ 

\newcommand{\bA} {\mathbf {A}}
\newcommand{\bB} {\mathbf {B}}
\newcommand{\bS} {\mathbf {S}}
\newcommand{\bV} {\mathbf {V}}
\newcommand{\bW} {\mathbf {W}}

\newcommand{\bU} {\mathbf {U}}

\newcommand{\bX} {\mathbf {X}}
\newcommand{\bx} {\mathbf {x}}
\newcommand{\bZ} {\mathbf {Z}}
\newcommand{\bbeta} {\mbox{\boldmath $\beta$}}

\newcommand{\bOmega} {\mbox{\boldmath $\Omega$}}
\newcommand{\bPsi} {\mbox{\boldmath $\Psi$}}
\newcommand{\bSigma} {\mbox{\boldmath $\Sigma$}}
\newcommand{\bzero} {\mbox{\boldmath $0$}}

\def\T{{ \mathrm{\scriptscriptstyle T} }}

\newcommand{\revise}{\textcolor{black}}

\begin{document}
	\maketitle
	
\begin{abstract}
An inference procedure is proposed to provide consistent estimators of parameters in a modal regression model with a  covariate prone to measurement error. A score-based diagnostic tool exploiting parametric bootstrap is developed to assess adequacy of parametric assumptions imposed on the regression model. The proposed estimation method and diagnostic tool are applied to synthetic data generated from simulation experiments and data from real-world applications to demonstrate their implementation and performance. These empirical examples illustrate the importance of adequately accounting for measurement error in the error-prone covariate when inferring the association between a response and covariates based on a modal regression model that is especially suitable for skewed and heavy-tailed response data.
\end{abstract}


\section{Introduction}
\label{sec:intro}

The mean, median, and mode are three  widely used measures of central tendency of data. The mode can be a more informative and sensible central tendency measure than the other two for data arising from distributions that are heavy-tailed and skewed. This very virtue of mode and  the ubiquity of heavy-tailed and skewed data in biology, sociology, economics, and many other fields of study have recently revived data scientists' interest in regression methodology focusing on the conditional mode of a response \citep{Chacn2020}. 

While there exists an extensive literature on regression models that relate the mean or the median of a response variable $Y$ to covariates \revise{$\bX$}, there are much less work on regression models tailored for the conditional mode of $Y$ given $\bX$ \citep{Sager1982, Lee1989, Lee1993}. Among the limited existing modal regression methods, \revise{the majority of them are in the semi-/non-parametric framework \citep{Yao2013, Chen2016, ota2019quantile, wang2019modal, kemp2020dynamic,zhang2021bootstrap, ullah2022nonlinear, xiang2022nonparametric}}, which typically suffer from low statistical efficiency when \revise{compared} with their parametric counterparts. One reality that discourages use of parametric models for inferring the mode is that very few named distributions that allow asymmetry can be conveniently formulated as distribution families indexed by the mode along with other parameters. Among the few groups of authors who considered parametric modal regression models, \citet[][Chapter 3]{aristodemou2014new} assumed a gamma distribution for a non-negative response with a covariate-dependent mode; \citet{bourguignon2020parametric} followed a similar model construction while also allowing a covariate-dependent precision parameter for the gamma distribution. Focusing on bounded response data, \citet{zhouhuang2020} proposed two modal regression models, one based on a beta distribution and the other based on a generalized biparabolic distribution for the response given covariates. In all three aforementioned works, frequentist likelihood-based methods are developed to infer model parameters. Most recently, \citet{zhou2022bayesian} unified the mean regression and modal regression in a Bayesian framework by reparameterizing a four-parameter beta distribution with an unknown support so that the mean or the mode of $Y$ depends on $\bX$. Earlier works on Bayesian modal regression, including parametric and nonparametric methods, can also be found in \citet[][Chapter 2]{aristodemou2014new}.

All the above works on modal regression assume that covariates are measured precisely. Data analysts in many disciplines are well aware that, among all variables of interest, some of them often cannot be measured precisely due to inaccurate measuring devices or human error in data collection. Some variables are in principle inaccessible and only some surrogates of them can be measured. For example, one's long-term blood pressure is an important biomarker associated with one's heart health, yet it cannot be directly measured. Instead, measurable surrogates of it are blood pressure readings collected during a doctor's visit, which can be viewed as error-contaminated versions of one's long-term blood pressure. It has also been well-understood that ignoring covariates measurement error in mean regression or quantile regression usually lead to misleading inference results. There exists a large collection of works on mean regression methodology accounting for measurement error \citep{Carroll2006, fuller2009measurement, Buonaccorsi2010, Yi2017}, and also some works in quantile regression to address this complication \citep{he2000quantile, wei2009quantile, wang2012corrected}. Modal regression methodology that address this issue only emerged recently, including those developed by \citet{Zhou2016}, \cite{lihuang2019}, and \cite{Shi2021}, all of which opted for a nonparametric model for the error term in the primary regression model. There is a lack of methodology to account for error-prone covariates in parametric modal regression, and our study presented in this article fills the void.

In preparation for proposing a method to account for measurement error in covariates that is applicable to any parametric modal regression models, we first formulate the measurement error model and discuss complications unique to modal regression models in Section~\ref{sec:datamodel}. For concreteness, we then focus on the beta modal regression model for a response supported on [0, 1] with an error-prone covariate, and propose consistent estimation methods to infer model parameters that account for measurement error in Section \ref{sec:estimation}. A model diagnostic method is developed to detect model misspecifications when adopting the beta modal regression model in a given application in Section~\ref{sec:diagnos}. Simulation studies are reported in Section~\ref{sec:simulation} to demonstrate the performance of the estimation and diagnostics methods. We apply the proposed modal regression method accounting for covariate measurement error to data sets arising from two real-life studies in Section \ref{sec:real}, where we also discuss revisions of the method to adapt to more general settings. Section \ref{sec:discussion} gives concluding remarks and future research directions.

\section{Data and Model}
\label{sec:datamodel}

\subsection{Observed data}

Suppose that, given $p$ covariates in $\bX=(X_1, \ldots, X_p)^\T$, $Y$ follows a unimodal distribution specified by the probability density function (pdf), $f_{\hbox {\tiny $Y|\bX$}}(y|\bx)$. Denote by $\theta(\bx)$ the mode of $Y$ given $\bX=\bx$. In modal regression without measurement error, one infers $\theta(\bx)$ based on a random sample of size $n$ from the joint distribution of $(Y,\bX)$, $\{(Y_j, \bX_j)\}_{j=1}^n$, where $\bX_j=(X_{1,j},\ldots, X_{p,j})^\T$. Now suppose that a covariate in $\bX$, say, $X_1$, is prone to measurement error, and a surrogate $W$ is observed instead of $X_1$, with $n_j$ replicate measures of $X_{1,j}$ in $\widetilde W_j=\{W_{j,k}\}_{k=1}^{n_j}$,  for $j=1, \ldots, n$. In this study, we assume that $W_{j,k}$ relates to $X_{1,j}$ via an additive measurement error model,  
\begin{equation}
	W_{j,k}=X_{1,j}+U_{j,k}, \mbox{ for $j=1, \ldots, n$ and $k=1, \ldots, n_j$,} \label{eq:wxu}
\end{equation}
where $\{U_{j,k}, \, k=1, \ldots, n_j\}_{j=1}^n$ are independent and identically distributed (i.i.d.) mean-zero measurement error, which are independent of $\{(Y_j, \bX_j)\}_{j=1}^n$ to guarantee nondifferential measurement error as considered in the classical measurement error models \citep[][Section 2.5]{Carroll2006}. 

In a naive univariate modal regression analysis using the surrogate data, one treats $W$ as if it were $X=X_1$, and equivalently, views the conditional pdf of $Y$ given $W=w$, $f_{\hbox {\tiny $Y|W$}}(y|w)$, the same as $f_{\hbox {\tiny $Y|X$}}(y|w)$. As a result, naive modal regression analysis essentially infers the mode of $f_{\hbox {\tiny $Y|W$}}(y|w)$ instead of $\theta(\cdot)$. In the context of univariate mean regression models not limited to linear regression, the attenuation effect of measurement error on covariate effect estimation is often noted in the literature \citep{Carroll2006, Buonaccorsi2010}, which causes the estimated covariate effect of a truly influential covariate to be pulled towards zero. Naive modal regression can suffer the same attenuation effect. For instance, if the mean and the mode of $f_{\hbox {\tiny $Y|X$}}(y|x)$ differ by a quantity that does not depend on covariates, such as for a Gumbel distribution that depends on a covariate $X$ only via the mode but not via the scale parameter, then the impact of measurement error on naive inference for the conditional mean mostly carries over to naive inference for $\theta(x)$. In other model settings where the conditional mean and mode of $Y$ differ by a quantity that does depend on the error-prone covariate, the effect of measurement error on naive modal regression demands investigation on a case-by-case basis. Even before conducting such investigation, a more fundamental question needs to be addressed, that is whether or not naive modal regression is meaningful, since unimodality of $f_{\hbox {\tiny $Y|X$}}(y|x)$ does not guarantee unimodality of $f_{\hbox {\tiny $Y|W$}}(y|w)$. Indeed, there is an extra layer of complication in modal regression with an error-prone covariate that does not exist in mean regression since, if the mean of $Y$ given $X$, $\mu(X)$, is well defined, then the mean of $Y$ given $W$ is $E\{\mu(X)|W\}$, which is also well defined in most settings of practical interest. Because of this additional complication, correcting naive inference to account for measurement error in modal regression  is more challenging than the counterpart task in mean regression. For example, a strategy that can be easy to implement in  mean regression is to correct the bias in a naive estimator of a parameter to produce an improved estimator accounting for measurement error \citep[][Section 3.4]{Carroll2006}. This idea of de-biasing naive estimation may not be a sensible approach now with the existence of a naive mode function in question.  

\subsection{Regression model}

We propose to account for measurement error when inferring parameters in a modal regression model by exploiting the idea of corrected scores. In particular, we focus on modeling a bounded response $Y$, which is commonly encountered in practice, such as test scores, disease prevalence, and the fraction of household income spent on food. Any bounded response with a known support can be scaled to be supported on the unit interval [0, 1]. Beta distribution is a parametric family that encompasses various shapes of distributions supported on [0, 1], and thus serves as a relatively flexible basis for building a regression model for such responses. For a random variable $V$ that follows a beta distribution with shape parameters $\alpha_1, \alpha_2>0$, i.e., $V\sim \mbox{beta}(\alpha_1, \alpha_2)$, its density function is,  
\[f(v;\alpha_1, \alpha_2) = \frac{\Gamma(\alpha_1+\alpha_2)}{\Gamma(\alpha_1) \Gamma(\alpha_2)}v^{\alpha_1-1}(1-v)^{\alpha_2-1}, \mbox{ for $0<v<1$,} \]
where \(\Gamma(\cdot)\) is the Gamma function. When \(\alpha_1, \, \alpha_2>1\), this distribution has a unique mode given by \(\theta = (\alpha_1-1)/(\alpha_1+\alpha_2-2)\). To prepare for modal regression, we reparameterize the beta
distribution by setting \(\alpha_1 = 1 + m\theta\) and \(\alpha_2 = 1+m(1-\theta)\), where $m>0$ plays the role of a precision parameter, with a larger value of $m$ leading to a smaller variance of the distribution \revise{\citep{zhouhuang2020}}. A similar parameterization of the beta distribution was used in \citet{chen1999beta} to formulate the beta kernel in kernel density estimators, and also in \citet{bagnato2013finite} to construct beta mixture distributions. In both earlier works, the beta family is indexed by $\theta$ and a dispersion parameter equal to the reciprocal of $m$. The parameterization of beta distributions used in our study is also in line with the one in \citet[][see Equation (6.6)]{kruschke2015doing}, except for that a concentration parameter equal to our $m$ plus 2 is used in place of our precision parameter there. Despite these small differences, all aforementioned parameterizations hightlight the mode as the location parameter, with the original shape parameters $\alpha_1$ and $\alpha_2$ specified by the mode and a precision/concentration/dispersion parameter that is of secondary interest in drawing inference. By construction, as long as the mode $\theta\in (0, 1)$ exists, which we assume throughout the study, we have \(\alpha_1, \alpha_2 > 1\) following our parameterization. 

With a beta distribution family indexed by $(\theta, m)$ formulated, a beta modal regression model follows by introducing covariates-dependent mode of $Y$, $\theta{(\mathbf{X})}=g(\bbeta^\T\tilde \bX)$, 
where $\tilde \bX=(1, \bX^\T)^\T$, $\bbeta=(\beta_0, \beta_1, \ldots, \beta_p)^\T$ with $\beta_0$ being the intercept and $\beta_1, \ldots, \beta_p$ representing covariate effects associated with the $p$ covariates in $\bX$, and \(g(\cdot)\) is a user-specified link function, such as logit, probit, log-log, and complementary log-log. Now a modal regression model for $Y$ is fully specified by the following conditional distribution of $Y$ given $\bX$, 
\begin{equation}
	Y|\mathbf{X} \sim \text{beta}(1+m\theta(\mathbf{X}), \, 1+m\{1-\theta(\mathbf{X})\}). \label{eq:yx}
\end{equation}
Combining (\ref{eq:yx}) with (\ref{eq:wxu}) completes the specification of a modal regression model for a response $Y$ supported on [0, 1] and covariates $\bX=(X_1, \ldots, X_p)^\T$, with $X_1$ subject to additive nondifferential measurement error. The focal point of inference lies in parameters involved in the primary regression model in (\ref{eq:yx}), $\bOmega=(\bbeta^\T, m)^\T$. Parameters appearing in (\ref{eq:wxu}) are of secondary interest but required to specify the measurement error distribution. 

\section{Parameter estimation}\label{sec:estimation}

\subsection{Maximum likelihood estimation}

In the absence of measurement error, one may carry out maximum likelihood estimation of $\bOmega$ straightforwardly by  solving the normal score equations for $\bOmega$. More specifically, the log-likelihood of error-free data,  $\mathcal{D}=\{(Y_j, \bX_j)\}_{j=1}^n$, is 
\begin{equation}
	\begin{aligned}
		\ell(\mathbf{\Omega} ; \mathcal{D}) = &\
		\sum_{j=1}^n \ell(\bOmega; Y_j, \bX_j)\\
		= &\ n \log \Gamma(2+m)-\sum_{j=1}^{n} \log \left(\Gamma(1+m \theta\left(\mathbf{X}_j\right)) \Gamma(1+m\left\{1-\theta\left(\mathbf{X}_j\right)\right\})\right) \\
		& +m \sum_{j=1}^{n}\left[\theta\left(\mathbf{X}_j\right) \log Y_j+\left\{1-\theta\left(\mathbf{X}_j\right)\right\} \log \left(1-Y_j\right)\right].
	\end{aligned}
	\label{eq:betaloglikelihood}
\end{equation}
Differentiating (\ref{eq:betaloglikelihood}) with respect to $\bOmega$ leads to the score equations, $\sum_{j=1}^n 
\bPsi_0(\bOmega; Y_j, \bX_j) =\bzero$, where the score vector evaluated at the $j$-th data point, $\bPsi_0(\bOmega; Y_j, \bX_j)$, consists of the following scores, for $j=1, \ldots, n$,
\begin{align}
	\frac{\partial \ell(\boldsymbol{\Omega}; Y_j, \bX_j)}{\partial \bbeta} = & \ \left\{-m \psi(1+m \theta(\mathbf{X}_j))+m \psi(1+m\{1-\theta(\mathbf{X}_j)\})+m \log \left(\frac{Y_j}{1-Y_j}\right)\right\}\nonumber \\ 
	&\ \times g'(\bbeta^\T \tilde \bX_j) \tilde\bX_j, \label{eq:scorebeta}\\
	\frac{\partial \ell(\boldsymbol{\Omega}; Y_j, \bX_j)}{\partial m}= & \ \psi(2+m)-\theta(\mathbf{X}_j) \psi(1+m \theta(\mathbf{X}_j))-\{1-\theta(\mathbf{X}_j)\} \psi(1+m\{1-\theta(\mathbf{X}_j)\}) \nonumber \\
	&\ +\theta(\mathbf{X}_j) \log Y_j+\{1-\theta(\mathbf{X}_j)\} \log (1-Y_j), \label{eq:scorem} 
\end{align}
where $\psi(t)=(d/dt)\log \Gamma(t)$ is the digamma function and $g'(t)=(d/dt)g(t)$. 

\subsection{Monte-Carlo corrected scores}
\label{sec:CSE}

In the presence of measurement error, a naive estimator of $\bOmega$ solves the naive score equations resulting from replacing $X_{1,j}$ with $\overline W_j=n_j^{-1}\sum_{k=1}^{n_j} W_{j,k}$ in (\ref{eq:scorebeta}) and (\ref{eq:scorem}), for $j=1, \ldots, n$. As pointed out earlier and also evidenced in simulation study to be presented later, this naive treatment typically results in misleading inference for $\bOmega$. We propose to follow the idea of the corrected score method \citep{Nakamura1990} and revise the naive scores to obtain estimating equations that adequately account for measurement error. The thrust of the corrected score method is to use the observed error-prone data, $\mathcal{D}^*=\{(Y_j, \widetilde W_j, \, \bX_{-1,j})\}_{j=1}^n$ with $\widetilde W_j=\{W_{j,k}\}_{k=1}^{n_j}$ and $\bX_{-1,j}=(X_{2,j}, \ldots, X_{p,j})^\T$, to construct unbiased estimators of the above normal scores. In this vein of thinking, one treats $\{X_{1,j}\}_{j=1}^n$ as unknown parameters instead of realizations of a random variable, and thus one takes on the functional point of view as opposed to the structural viewpoint of measurement error models where a distribution for $X_1$ is assumed \citep[][Section 2.1]{Carroll2006}.

We begin with applying the Monte-Carlo-amenable method proposed by \citet{Stefanski2005}, a method originating from the idea described in \citet{stefanski1989unbiased}. More specifically, we construct a score, $\bPsi(\bOmega; Y_j, \widetilde W_j, \bX_{-1,j})$, that satisfies $E\{\bPsi(\bOmega;Y_j, \widetilde W_j, \bX_{-1,j})|Y_j, \bX_j\}=\bPsi_0(\bOmega; Y_j, \bX_j)$, for $j=1, \ldots, n$. This particular method is especially suitable for settings with a univariate error-prone covariate subject to normal measurement error $U$. We will address violation of the normality assumption on $U$ in Section~\ref{sec:estimation}, and describe revisions of the method to adapt to settings  with multiple error-prone covariates in Section~\ref{sec:real}. As shown in   \citet[][Theorem 1]{Stefanski2005}, the minimum variance unbiased estimator of $\bPsi_0(\bOmega; Y_j, \bX_j)$ is given by 
\begin{equation}
	\bPsi(\bOmega; Y_j, \widetilde W_j, \bX_{-1,j})  = E\left\{ \left. \bPsi_0\left( \bOmega; Y_j, \overline W_j +i \sqrt{\frac{(n_j-1)S_j^2}{n_j}} T, \bX_{-1,j}\right) \right\vert Y_j,  \overline W_j, S_j^2, \bX_{-1,j}\right\},
	\label{eq:theorem1}
\end{equation}
where $i$ is the imaginary unit, $S_j^2$ is the sample variance of $\widetilde W_j=\{W_{j,k}\}_{k=1}^{n_j}$, and $T=Z_1/(\sum_{k=1}^{n_j-1}Z^2_k)^{1/2}$ is independent of all observed data, in which $Z_1, \ldots, Z_{n_j-1}$ are independent standard normal random variables. The estimator of $\bPsi_0(\bOmega; Y_j, \bX_j)$ in (\ref{eq:theorem1}) originates from a jackknife exact-extrapolant estimator constructed for the purpose of estimating a function of the mean of a normal distribution based on a random sample from the distribution. In the context of (\ref{eq:theorem1}), this random sample is $\widetilde{W}_j$ from $N(X_{1,j}, \sigma_u^2)$, where $\sigma_u^2$ is the measurement error variance, i.e., assuming $U\sim N(0, \sigma_u^2)$ in (\ref{eq:wxu}), and the function of the normal mean $X_{1,j}$ is $\bPsi_0(\bOmega; Y_j, X_{1,j}, \bX_{-1,j})$. The expectation in (\ref{eq:theorem1})  cannot be derived in closed form. But since the only quantity viewed as random when deriving this conditional expectation is $T$ that is independent of observed data, one can estimate this expectation unbiasedly via an empirical mean based on simulated random samples of $T$. Moreover, as shown in \citet{Stefanski2005}, even though (\ref{eq:theorem1}) is  complex-valued by construction, the expectation of its imaginary part is zero as long as $\bPsi_0(\bOmega; Y_j, X_{1,j}, \bX_{-1,j})$ is infinitely differentiable with respect to $X_{1,j}$, which is guaranteed in our case by choosing a link function $g(t)$ that is infinitely differentiable. Hence, using the real part of the empirical version of (\ref{eq:theorem1}) suffices for constructing an unbiased estimator of $\bPsi_0(\bOmega; Y_j, \bX_j)$. This leads to the following corrected score based on a simulated random sample of $T$ of size $B$, $\widetilde T_j=\{T_{j,b}\}_{b=1}^B$, for $j=1,\ldots, n$,
\begin{equation}
	\bPsi(\bOmega; Y_j, \widetilde W_j, \widetilde T_j, \bX_{-1,j})
	=\frac{1}{B} \sum_{b=1}^{B} \mbox{Re}\left\{\bPsi_0\left(\bOmega; Y_j, \overline W_j +i \sqrt{\frac{(n_j-1)S_j^2}{n_j}} T_{j,b}, \bX_{-1,j} \right)\right\},
	\label{eq:montecarlo}
\end{equation}
where $\mbox{Re}(t)$ denotes the real part of a complex-valued $t$. 

One now can solve the following system of $p+2$ equations based on the corrected score in (\ref{eq:montecarlo}),
\begin{equation}
	\sum_{j=1}^n \bPsi(\bOmega; Y_j, \widetilde W_j, \widetilde T_j, \bX_{-1,j})=\bzero, \label{eq:complexeq}
\end{equation} 
for $\bOmega$ to obtain a consistent estimator $\hat \bOmega$, where $\widetilde T_1, \ldots, \widetilde T_n$ are independent. Solving (\ref{eq:complexeq}) for $\bOmega$ is equivalent to solving an optimization problem, that is,
\begin{equation}
	\hat{\mathbf{\Omega}} = \underset{\mathbf{\Omega} \in \mathbb{R}^{p+1}\times \mathbb{R}^+}{\arg \min } \left\{\sum_{j=1}^n\bPsi(\bOmega; Y_j, \widetilde W_j, \widetilde T_j, \bX_{-1,j})\right\}^\T
	\left\{\sum_{j=1}^n\bPsi(\bOmega; Y_j, \widetilde W_j, \widetilde T_j, \bX_{-1,j})\right\}.
	\label{eq:mestimator}
\end{equation}
The equivalence between  (\ref{eq:mestimator}) and the solution to (\ref{eq:complexeq}) is obvious when there exists a unique solution to (\ref{eq:complexeq}). An added benefit of dealing with an optimization problem is more appreciated in the presence of model misspecification that can potentially lead to non-existence of a solution to (\ref{eq:complexeq}), yet (\ref{eq:mestimator}) may still be well-defined with meaningful statistical interpretations according to \citet{white1982maximum}.  

\subsection{Monte-Carlo corrected log-likelihood}
\label{sec:MCCL}
To this end, estimating $\bOmega$ appears to be a straightforward optimization problem. But the numerical procedure to obtain (\ref{eq:mestimator}) requires evaluating $p+2$ scores at each iteration, which can be cumbersome and very demanding on the computer memory and central processing unit, especially due to the Monte Carlo nature of the score in (\ref{eq:montecarlo}) that involves computing a vector-valued score $B$ times. Viewing the quadratic form in (\ref{eq:mestimator}) as an objective function that accounts for measurement error, we propose to use a different objective function that also takes measurement error into account and is computationally less cumbersome to optimize. This new objective function is obtained by correcting the naive log-likelihood function $\ell(\bOmega; Y_j, \overline W_j, \bX_{-1,j})$ that is the summand of (\ref{eq:betaloglikelihood}) with $X_{1,j}$ evaluated at $\overline W_j$, for $j=1, \ldots, n$. Similar to the construction of the corrected score in (\ref{eq:montecarlo}) based on the naive score, the new objective function based on the naive log-likelihood evaluated at the $j$-th observed data point is
\begin{align}
	\tilde\ell(\bOmega; Y_j, \widetilde W_j, \widetilde T_j, \bX_{-1,j})  & = \frac{1}{B} \sum_{b=1}^{B} \mbox{Re}\left\{\ell\left(\bOmega; Y_j, \overline W_j +i \sqrt{\frac{(n_j-1)S_j^2}{n_j}} T_{j,b}, \bX_{-1,j} \right)\right\}, \label{eq:correctedloglkh}
\end{align}
which satisfies $E\{\tilde\ell(\bOmega; Y_j, \widetilde W_j, \widetilde T_j, \bX_{-1,j})|Y_j, \bX_j\}=\ell(\bOmega; Y_j, \bX_j)$,  for $j=1, \ldots, n$. We then define an estimator of $\bOmega$ as 
\begin{equation}
	\hat{\mathbf{\Omega}} = \underset{\mathbf{\Omega} \in \mathbb{R}^{p+1}\times \mathbb{R}^+}{\arg \max } \sum_{j=1}^{n} \tilde\ell(\bOmega; Y_j, \widetilde W_j, \widetilde T_j, \bX_{-1,j}),
	\label{eq:mestimatorfast}
\end{equation}
which only requires repeated evaluation of a scalar function in (\ref{eq:correctedloglkh}) at each iteration of an optimization algorithm. In simulation studies (not presented in this article) where we estimate $\bOmega$ using these two routes of optimization according to (\ref{eq:mestimator}) and (\ref{eq:mestimatorfast}), we obtain very similar estimates of $\bOmega$, with the former route more computationally demanding than the latter. The numerical similarity of (\ref{eq:mestimator}) and (\ref{eq:mestimatorfast}) may be expected given the connection between the naive score and the naive log-likelihood, in addition to the equivalence between the solution to the normal score equation and the maximum likelihood estimator in the absence of measurement error. We refer to the estimator defined in (\ref{eq:mestimatorfast}) the Monte Carlo corrected log-likelihood estimator, or MCCL for short. 

Whether one follows the idea of correcting the naive scores or the route of correcting the naive log-likelihood to account for measurement error, our proposed estimation method falls in the general framework of $M$-estimation \citep[][Chapter 7]{boos2013essential}. As an $M$-estimator, the MCCL estimator $\hat \bOmega$ is a consistent estimator of $\bOmega$ that is asymptotically normal under regularity conditions 
stated in, for example, Theorem 7.2 in \citet{boos2013essential}. Moreover, motivated by its asymptotic variance of the sandwich form \citep[][Section 7.2.1]{boos2013essential}, the variance of $\hat\bOmega$ can be estimated by 
\begin{equation} 
	\bV(\mathcal{D}^*; \hat \bOmega)
	= \left\{\bA(\mathcal{D}^*; \hat \bOmega)\right\}^{-1}\bB(\mathcal{D}^*; \hat \bOmega)\left[\left\{\bA(\mathcal{D}^*; \hat \bOmega)\right\}^{-1}\right]^\T,
	\label{eq:sandwich}
\end{equation} 
where
\begin{align*}
	\bA(\mathcal{D}^*; \hat \bOmega) & = \left. \frac{1}{n}\sum_{j=1}^n \frac{\partial}{\partial \bOmega^\T} \bPsi(\bOmega; Y_j, \widetilde W_j, \widetilde T_j, \bX_{-1,j})\right\vert_{\bOmega=\hat \bOmega}, \\
	\bB(\mathcal{D}^*; \hat \bOmega) & =  \frac{1}{n}\sum_{j=1}^n \bPsi(\hat\bOmega; Y_j, \widetilde W_j, \widetilde T_j, \bX_{-1,j}) \left\{\bPsi(
	\hat\bOmega; Y_j, \widetilde W_j, \widetilde T_j, \bX_{-1,j})\right\}^\T .
\end{align*}

\section{Model diagnostics}
\label{sec:diagnos}
Even though we avoid specifying the true covariate distribution by adopting the functional viewpoint of measurement error models, the primary regression model in (\ref{eq:yx}) is fully parametric. This raises the concern of model misspecification and calls for model diagnostics tools. Model diagnostics based on error-prone data is more challenging than settings without measurement error. In particular, conventional residual-based diagnostics methods that require evaluating an estimated regression function, whether it is the conditional mean $\mu(\bX)$ in mean regression or the conditional mode $\theta(\bX)$ in modal regression, are no longer applicable now that a true covariate is unobserved. Another contribution of our study is an effective score-based diagnostic tool that circumvents this obstacle a traditional residual-based diagnostic method faces in the presence of measurement error. 

For the beta modal regression model without error in covariates, \citet{zhouhuang2020} propose a
score-based test statistic defined below for the purpose of model diagnostics, 
\begin{equation}
	Q(\hat{\mathbf{\Omega}}_0;\mathcal{D}) = \frac{n-2}{2(n-1)}\overline{\mathbf{S}}^\T\hat{\mathbf{\Sigma}}^{-1}\overline{\mathbf{S}},
	\label{eq:Qstatistics}
\end{equation}
where $\hat \bOmega_0$ is the maximum likelihood estimator of $\bOmega$, $\overline{\mathbf{S}} =n^{-1} \sum_{j=1}^{n} \mathbf{S}(\hat{\mathbf{\Omega}}_0; Y_j, \bX_j)$, and 
$\hat{\boldsymbol{\Sigma}}  =\{n(n-1)\}^{-1} \sum_{j=1}^{n}\{\mathbf{S}(\hat{\boldsymbol{\Omega}}_0; Y_j, \bX_j)-\overline{\mathbf{S}}\}\{\mathbf{S}(\hat{\boldsymbol{\Omega}}_0; Y_j, \bX_j)-\overline{\mathbf{S}}\}^\T$, in which, for $j=1, \ldots, n$,
\begin{align}
	\mathbf{S}(\boldsymbol{\Omega}; Y_j, \bX_j) & =
	\begin{bmatrix}
		\log Y_j-\psi(1+m \theta(\bX_j))+\psi(2+m) \\
		\displaystyle{Y_j \log Y_j-\frac{\{1+m \theta(\bX_j)\}\{\psi(2+m \theta(\bX_j))-\psi(3+m)\}}{2+m}}
	\end{bmatrix}\label{eq:S}
\end{align} 
is the score vector constructed by matching $\log V$ and $V\log V$ with their respective expectations for $V\sim \mbox{beta}(\alpha_1, \alpha_2)$, and thus $E\{\bS(\bOmega; Y_j, \bX_j)\}=\bzero$ in the absence of model misspecification. By construction, a larger value of the nonnegative $Q(\hat \bOmega_0; \mathcal{D})$ provides stronger evidence indicating model misspecification. A parametric bootstrap procedure is developed in \citet{zhouhuang2020} to estimate the null distribution of $Q(\hat \bOmega_0; \mathcal{D})$, from which one onbtains an estimated $p$-value for the test.

Returning to our beta modal regression model with error-in-covariate, we apply the idea of corrected score here to construct a counterpart of (\ref{eq:S}) to obtain a score accounting for measurement error whose mean is zero in the absence of model misspecification. This yields the corrected score evaluated at the $j$-th observed data point for model diagnostics, for $j=1, \ldots, n,$
\begin{equation}
	\tilde \bS(\bOmega; Y_j, \widetilde W_j, \widetilde T_j, \bX_{-1,j}) =   \frac{1}{B} \sum_{b=1}^{B} \mbox{Re}\left\{\bS\left(\bOmega; Y_j, \overline W_j +i \sqrt{\frac{(n_j-1)S_j^2}{n_j}} T_{j,b}, \bX_{-1,j} \right)\right\}.
	\label{eq:Stilde}
\end{equation}
The test statistic of the quadratic form denoted by $\tilde Q(\hat\bOmega; \mathcal{D}^*)$ that is parallel to  (\ref{eq:Qstatistics}) follows by using the MCCL estimator $\hat \bOmega$ instead of $\hat \bOmega_0$,  replacing $\overline \bS$ appearing in (\ref{eq:Qstatistics}) with $n^{-1}\sum_{j=1}^n \tilde \bS(\bOmega; Y_j, \widetilde W_j, \widetilde T_j, \bX_{-1,j})$, and revising $\hat \bSigma$ accordingly. But the next hurdle emerges, that is the design of a parametric bootstrap procedure for estimating the null distribution of $\tilde Q(\hat\bOmega; \mathcal{D}^*)$. Traditional parametric bootstrap in the regression setting, such as the procedure in \citet{zhouhuang2020}, involves generating response data from the primary regression model that  again requires evaluating an estimated regression function at the true covariates that are partly unobserved in the current context. We overcome this hurdle by ``estimating'' unobserved true covariate data, as implemented in the method of regression calibration \citep[Chapter 4,][]{Carroll2006} that takes on the structural viewpoint of measurement error models. Under the classical measurement error in (\ref{eq:wxu}), the best linear predictor of $X_{1,j}$ is $E(X_{1,j}|\overline W_j)=\mu_1+\lambda_j (\overline W_j-\mu_1)$, where $\mu_1=E(X_1)$ and $\lambda_j=n_j\sigma^2_1/\sigma^2_{\hbox {\tiny $W$}}$ is the reliability ratio associated with $\overline W_j$ \citep[][Section 3.2.1]{Carroll2006}, in which $\sigma^2_1$ and $\sigma^2_{\hbox {\tiny $W$}}$ denote the variance of $X_1$ and that of $W$, respectively. Replacing each unknown quantity in $E(X_{1,j}|\overline W_j)$ with its method-of-moments estimator yields an ``estimator" or prediction of $X_{1,j}$ given by  
\begin{equation}
	\hat X_{1,j}^*= \overline W+\hat \lambda (\overline W_j-\overline W), \text{ for $j=1,\ldots, n$,} \label{eq:Xhat}
\end{equation}
where $\overline W=n^{-1}\sum_{j=1}^n \overline W_j$ and $\hat \lambda=\hat \sigma^2_1/\hat \sigma^2_{\hbox {\tiny $W$}}$, in which $\hat\sigma^2_{\hbox {\tiny $W$}}$ is the sample variance of $(\overline W_1, \ldots, \overline W_n)$, $\hat \sigma^2_1=(\hat\sigma^2_{\hbox {\tiny $W$}}-\hat \sigma^2_u)_+$, and $\hat \sigma_u^2=n^{-1}\sum_{j=1}^n S_j^2/n_j$,
recalling that, for $j=1,\ldots, n$, $S_j^2$ is the sample variance of $(W_{j,1}, \ldots, W_{j,n_j})$ computed earlier  to evaluate the corrected score and the corrected log-likelihood. The idea of regression calibration is to regress $Y$ on the estimated covariate $\hat X^*_1$ defined by (\ref{eq:Xhat}) and $\bX_{-1}=(X_2, \ldots, X_p)^\T$ instead of regressing on $(W , \bX_{-1}^\T)^\T$ . Even though this idea often yields estimators of parameters in the primary regression model improved over naive estimators, \citet{modelbootstrap} noted that (\ref{eq:Xhat}) tends to underestimate the variability of the true covariate and thus can be problematic if used in a bootstrap procedure as we intend to. They then proposed to use 
\begin{equation}
	\hat X_{1,j}= \overline W+\hat \lambda^{1/2} (\overline W_j-\overline W), \text{  for $j=1, \ldots, n$,}\label{eq:Xhat2}
\end{equation}
as estimated covariate data instead so that these estimated covariate values have the mean and variance coinciding with method-of-moments estimates for the mean and variance of $X_1$. 

With this last hurdle resolved, we are in the position to present the detailed algorithm of the parametric bootstrap for estimating the $p$-value associated with $\tilde Q(\hat\bOmega; \mathcal{D}^*)$ based on $M$ bootstrap samples next.
\begin{enumerate}
	\item[]{\it Step 1}: 
	Fit the beta modal regression model with classical measurement error to $\mathcal{D}^*$ by applying the MCCL method in Section~\ref{sec:MCCL}. This gives the MCCL estimate $\hat \bOmega=(\hat \bbeta^\T, \hat m)^\T$.
	\item[]{\it Step 2}: 
	Compute the test statistic $\tilde Q(\hat\bOmega; \mathcal{D}^*)$.
	\item[]For $d=1,\ldots, M$, repeat {\it Steps 3--5},  
	\item[]{\it Step 3}: 
	For $j=1,\ldots, n$, generate \(Y^{(d)}_j\) from
	beta\((1+\hat{m} \hat \theta(\hat X_{1,j}, \bX_{-1,j}), \, 1+\hat{m}\{1-\hat\theta(\hat X_{1,j}, \bX_{-1,j})\})\), and generate
	\(W^{(d)}_{j,k} = \hat X_{1,j} + U^{(d)}_{j,k}\),
	for \(k = 1\), \ldots, \(n_j\), 
	where $\hat X_{1,j}$ is given by (\ref{eq:Xhat2}), and \(\{U^{(d)}_{j,k}\}_{k=1}^{n_j}\) are i.i.d. from $N(0, S_j^2)$. Let $\widetilde W_j^{(d)}=\{W^{(d)}_{j,k}\}_{k=1}^{n_j}$. This yields the $d$-th set of bootstrap data,
	\(\mathcal{D}^{(d)} = \{(Y^{(d)}_j, \, \widetilde W^{(d)}_j, \bX_{-1,j})\}_{j=1}^n\).
	\item[]{\it Step 4}: Fit the beta modal regression model with classic measurement error to $\mathcal{D}^{(d)}$, and obtain the MCCL estimate of $\bOmega$, denoted by 
	\(\hat{\boldsymbol{\Omega}}^{(d)}\).
	\item[]{\it Step 5}: 
	Compute the test statistic,
	\(\tilde Q(\hat{\boldsymbol{\Omega}}^{(d)}; \mathcal{D}^{(d)})\).
	\item[]{\it Step 6}: 
	Estimate the \(p\)-value by
	\(M^{-1}\sum_{d=1}^M I\left\{\tilde Q\left(\hat{\boldsymbol{\Omega}}^{(d)}; \mathcal{D}^{(d)}\right) > \tilde Q\left(\hat{\boldsymbol{\Omega}}; \mathcal{D}^*\right)\right\}\).
\end{enumerate}

\revise{In the absence of covariate measurement error where $\{X_{1,j}, \, j=1, \ldots, n\}$ are observed, the above algorithm (with $\hat X_{1,j}$ replaced by $X_{1,j}$ in {\it Step 3}) essentially follows the general guidelines of  bootstrap hypothesis testing as discussed in \citet{hall1991two}, \citet{davison1997bootstrap}, and \citet{martin2007bootstrap}. In particular, our targeted null hypothesis states that the response given true covariates follows a beta modal regression model; {\it Step 1} in our bootstrap algorithm aims to ``recover" the model consistent with the null, and response data obtained in {\it Step 3} are generated from the fitted null model and thus these response data reflect the null. This is precisely the first principle of model-based bootstrap for hypothesis testing: to generate bootstrap data that reflect the null. The unique challenge of bootstrap hypothesis testing in the presence of covariate measurement error is that true covariate values need to be estimated before generating response data. Unlike response data generation, which should reflect the null (that does not specify a distribution for the true covariate data), when ``recovering" true covariate values, one aims to recover certain structures of the design matrix in the absence of measurement error. We accomplish this goal by using $\{\hat X_{1,j}, \, j=1, \ldots, n\}$ in (\ref{eq:Xhat2}), which preserve certain structures of true covariate values in the sense that the first two moments of these estimated covariate values coincide  with the method-of-moment estimates for the first two moments of $\{X_{1,j}, \, j=1, \ldots, n\}$. The so-constructed estimated true covariate values are also used in \citet{thomas2011moment} to recover true covariate data. Even though it is unclear if there exists a better way to  recover error-free covariates data for the purpose of bootstrap hypothesis testing, \citet{buonaccorsi2016correcting} showed that this approach substantially outperforms two  obvious alternative methods: one is to use $\overline W_j$ to estimate $X_{1,j}$, the other is to use $\hat X_{1,j}^*$ in (\ref{eq:Xhat}). In our context,} empirical evidence from the simulation study presented in the next section suggest that the proposed bootstrap procedure can estimate the null distribution of $\tilde Q(\hat \bOmega; \mathcal{D}^*)$ accurately enough to preserve the right size of the test for model misspecification over a wide range of significance levels. 

\section{Simulation study}\label{sec:simulation}
We carry out simulation study to inspect finite sample performance of the proposed estimation method and the diagnostic method. The source code to reproduce results in this section is publicly available on the journal's web page.

\hypertarget{design-of-simulation-experiments}{%
	\subsection{Design of simulation
		experiments}\label{design-of-simulation-experiments}}
We generate data from each of the following four data generation processes.
\begin{enumerate}
	\item[(M1)]Generate response data according to (\ref{eq:yx}), with $m=3$, \(\theta(\mathbf{X})=1/\{1+\exp(-\beta_0-\beta_1X_1-\beta_2X_2)\}\), $\bbeta=(\beta_0, \beta_1, \beta_2)^\T=(0.25, 0.25, 0.25)^\T$, 
	\(X_2\sim \text{Bernoulli}(0.5)\), and  \(X_1|X_2\sim N(I(X_2=1)-I(X_2=0), \, 1)\),
	where $I(\cdot)$ is the indicator function. Contaminate data of $X_1$ according to (\ref{eq:wxu}) to generate $W_{j,k}$, for $j=1, \ldots, n$ and $k=1, 2, 3$, with \(U_{j,k} \sim N(0,\sigma^2_{u})\).
	\item[(M2)]Same as (M1) except for that $m=40$ and \(\theta(\mathbf{X})=1/\{1+\exp(-\beta_0-\beta_1X_1-\beta_2X_2-\beta_3 X_1^2)\}\), with $\bbeta=(\beta_0, \beta_1, \beta_2, \beta_3)^\T=(1, 1, 1, 1)^\T$.
	\item[(M3)]Same as (M1) except for that 
	\(\theta\left(\mathbf{X}\right)=\Phi\left(\beta_0+\beta_1X_1+\beta_2X_2\right)\) with $\bbeta=(\beta_0, \beta_1, \beta_2)^\T=(1, 1, 1)^\T$, where \(\Phi(\cdot)\) is the cumulative distribution function of $N(0, 1)$.
	\item[(M4)]
	Generate response data  $\{Y_j\}_{j=1}^n$ according to  $Y_j=(Y_j^*-Y^*_{(1)})/(Y^*_{(n)}-Y^*_{(1)})$, for $j=1, \ldots, n$, where $Y^*_{(1)}$ and $Y^*_{(n)}$ are the minimum and maximum order statistics of data $\{Y^*_j\}_{j=1}^n$, respectively, $Y^*_j\mid \mathbf{X}_j \sim \operatorname{Gumbel}(\theta(\mathbf{X}_j),\, \gamma^{-1}\{1-2\theta(\mathbf{X}_j)\}/(2+m))$,  
	in which $\theta(\mathbf{X}_j)<0.5$ is the mode formulated as that in (M1) with $\bbeta=(\beta_0, \beta_1, \beta_2)=(1, 1, 1)^\T$,  $\gamma^{-1} \{1-2\theta(\mathbf{X}_j)\}/(2+m)$ is the scale of the Gumbel distribution, and $\gamma$ stands for the Euler–Mascheroni constant.
\end{enumerate}

Despite the data generation process used to generate a particular data set, we always assume a beta modal regression model with $\theta(\bX)$ specified as that in (M1) when carrying out modal regression analysis of $Y$ on $\bX=(X_1, X_2)^\T$. By so doing, the design in (M1) allows us to  monitor point estimation in the absence of model misspecification, and the latter three designs can be used to study operating characteristics of the proposed model diagnostic method in the presence of different sources of model misspecification. In particular, fitting the assumed model to data generated according to (M2) creates a scenario where one misspecifies the linear predictor in the regression function. When data are generated from (M3), the assumed model has a wrong link function. Finally, fitting the assumed model to data from (M4) gives rise to the most severe model misspecification in the sense that the true distribution of $Y$ given $\bX$ is outside of the beta family. 

\hypertarget{performance-of-estimators}{%
	\subsection{Performance of point estimation}\label{performance-of-estimators}}
Besides assessing the quality of the MCCL estimator of $\bOmega$ in comparison with the naive maximum likelihood estimator, we aim at addressing the following three issues of point estimation in the simulation study: (i) the impact of having an error-free covariate along with an error-prone covariate on covariate effects estimation; (ii) the quality of the variance estimation based on (\ref{eq:sandwich}); (iii) the robustness of the MCCL estimator to the normality assumption on $U$. We bring up the third issue because the corrected score method is developed under the assumption of normal measurement error. Due to our focus on covariate effects estimation in the presence of an error-prone covariate in a modal regression model for a bounded response, none of the existing modal regression methods accounting for measurement error referenced in Section~\ref{sec:intro} serves as a sensible competing method in the current simulation study (e.g., there is no covariate effect parameters $\bbeta$ in a nonparametric modal regression model) .

Based on data generated according to (M1) with $\sigma^2_u=0.6, 1.2$, we obtain the MCCL estimate of $\bOmega$ using \revise{$B=100, 200$} and the naive maximum likelihood estimate that ignores measurement error in $X_1$. Table~\ref{tab:tab1} provides the median of MCCL estimates $\hat \bOmega$ and the median of naive estimates across 1000 Monte Carlo replicates at each of the two sample sizes $n=100, 200$. In contrast to the naive estimates that exhibit bias that do not diminish as the sample size increases, the MCCL estimates are much improved despite the severity of error contamination in $X_1$. \revise{And raising $B$ from 100 to 200 provides negligible improvement in the qualitiy of MCCL estimates. We thus set $B=100$ in the remaining empirical study and only show results corresponding to this default choice of $B$ in the sequel.} Not surprisingly, the MCCL estimator corrects the bias of the naive estimator at the price of an inflation in variation.  
\begin{table}[h]
	\caption{\label{tab:tab1}Medians of MCCL estimates and medians of naive estimates across $1000$
		Monte Carlo replicates generated according to (M1). The number in parentheses following each median is the interquartile range of the 1000 realizations of an estimator.}
	\centering
	\begin{tabular}[t]{*{2}{l}*{4}{c}}
		\toprule
		&  & $\beta_0$ & $\beta_1$ & $\beta_2$ & $\log m$ \\
		\midrule
		\multicolumn{2}{c}{ } & \multicolumn{4}{c}{$\sigma^2_u=0.6$} \\
		\cmidrule(l{3pt}r{3pt}){3-6} 
		& $\text{MCCL}_{B=100}$ & 0.23 (0.34) & 0.24 (0.22) & 0.26 (0.59) & 1.18 (0.30) \\
		& \revise{$\text{MCCL}_{B=200}$} & 0.23 (0.35) & 0.24 (0.22) & 0.26 (0.59) & 1.18 (0.30) \\
		\multirow{-3}{*}{\raggedright\arraybackslash $n=100$} & Naive   & 0.19 (0.31) & 0.20 (0.18) & 0.35 (0.55) & 1.16 (0.29) \\
		\cmidrule{1-6}
		& $\text{MCCL}_{B=100}$ & 0.24 (0.23) & 0.25 (0.15) & 0.26 (0.40) & 1.14 (0.22) \\
		& \revise{$\text{MCCL}_{B=200}$} & 0.24 (0.23) & 0.25 (0.15) & 0.26 (0.40) & 1.14 (0.22) \\
		\multirow{-3}{*}{\raggedright\arraybackslash $n=200$} & Naive   & 0.20 (0.22) & 0.20 (0.13) & 0.34 (0.37) & 1.13 (0.21) \\
		\hline
		\multicolumn{2}{c}{ } & \multicolumn{4}{c}{$\sigma^2_u=1.2$} \\
		\cmidrule(l{3pt}r{3pt}){3-6} 
		& $\text{MCCL}_{B=100}$ & 0.24 (0.34) & 0.24 (0.24) & 0.27 (0.65) & 1.18 (0.30)\\
		& \revise{$\text{MCCL}_{B=200}$} & 0.24 (0.35) & 0.24 (0.24) & 0.26 (0.66) & 1.18 (0.30)\\
		\multirow{-3}{*}{\raggedright\arraybackslash $n=100$} & Naive  & 0.17 (0.31) & 0.17 (0.17) & 0.41 (0.54) & 1.16 (0.29)\\
		\cmidrule{1-6}
		& $\text{MCCL}_{B=100}$ & 0.25 (0.25) & 0.25 (0.18) & 0.25 (0.43) & 1.14 (0.21)\\
		& \revise{$\text{MCCL}_{B=200}$} & 0.25 (0.25) & 0.25 (0.18) & 0.26 (0.43) & 1.14 (0.21)\\
		\multirow{-3}{*}{\raggedright\arraybackslash $n=200$} & Naive   &  0.17 (0.21) & 0.17 (0.12) & 0.41 (0.36) & 1.12 (0.21)\\
		\bottomrule
	\end{tabular}
\end{table}

\begin{figure}[h!]
	\centering
	\includegraphics{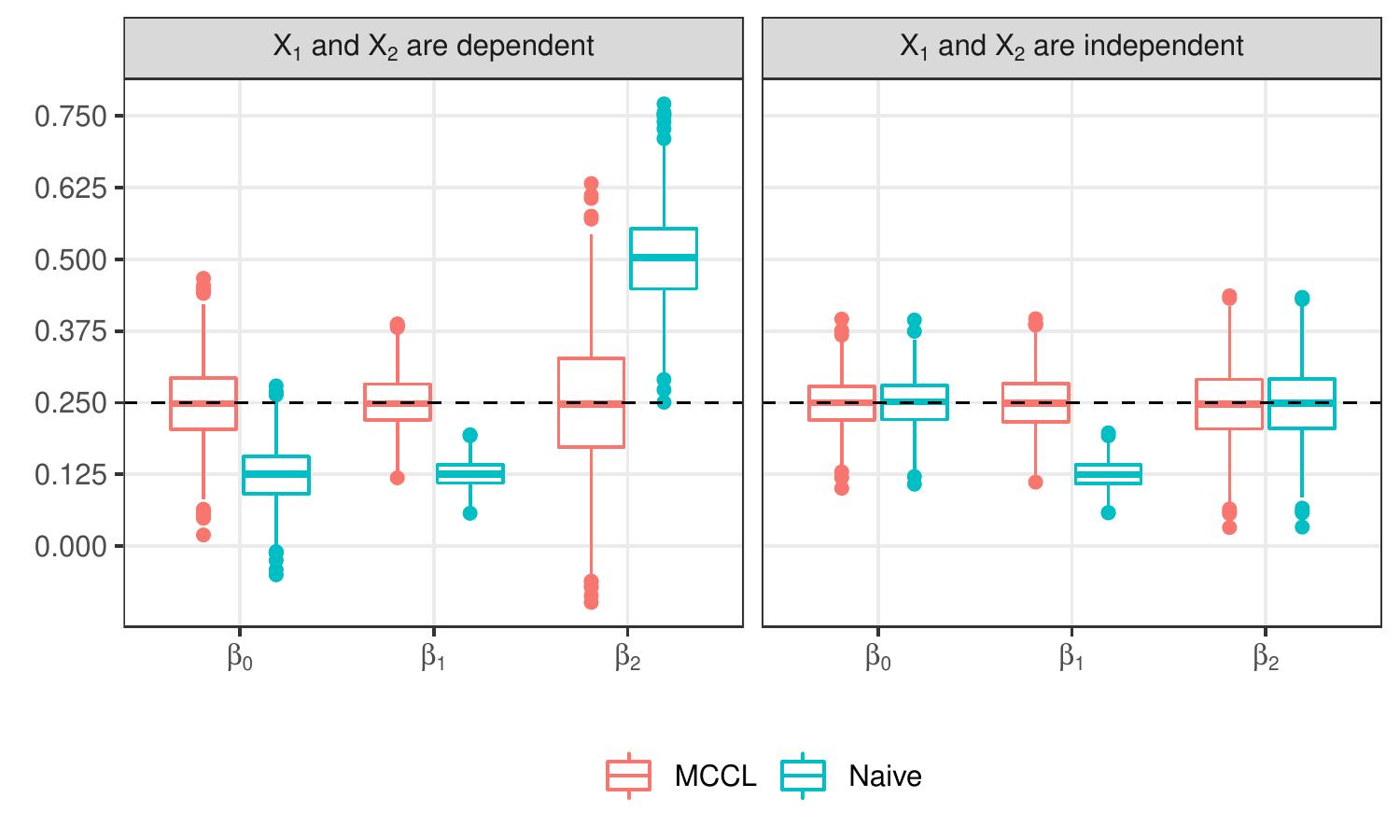}
	\caption{Boxplots of regression coefficients estimates under (M1) with $X_1$ and $X_2$ dependent (left panel) and those under a revised version of (M1) with $X_1$ and $X_2$ independent (right panel). The two boxes associated with  each parameter correspond to two estimators (from left to right): the MCCL estimator (red box) and the naive estimator (cyan box). } 
	\label{fig:fig1}
\end{figure}

The attenuation effect of measurement error on the naive covariate effect estimation for $X_1$ is evident in Table~\ref{tab:tab1}. In contrast, the covariate effect estimation for the error-free covariate $X_2$ is noticeably overestimated by the naive method. One may wonder if the observed opposite directions in the bias of naive estimation of two covariates effects persists when the two covariates are independent. This relates to the first issue brought up above. To address this issue, we revise the data generating process in (M1) in that $X_1\sim N(0, 1)$. 
Figure \ref{fig:fig1} includes boxplots of two sets of regression coefficients estimates, including the MCCL estimates and the naive estimates, under (M1) where $X_1$ and $X_2$ are dependent (see the left panel in Figure \ref{fig:fig1}) and under the revised (M1) with $X_1$ and $X_2$ independent (see the right panel in Figure \ref{fig:fig1}). Here, we set $n=2000$ for each of 1000 Monte Carlo replicates. Interestingly, when $X_2$ is independent of the error-prone covariate $X_1$, naive estimation for the covariate effect of $X_2$ does not appear to be affected by measurement error. Regardless, the attenuation in the estimated covariate effect for $X_1$ remains. 

Table \ref{tab:tab2} presents the average of standard deviation estimation of each parameter in $\bOmega$ based on (\ref{eq:sandwich}) across 1000 Monte Carlo replicates from (M1) with $n=200$. The Monte Carlo standard deviation of each parameter estimate in $\bOmega$ is used as a reference/gold standard in this table. The proximity of the standard deviation estimate with the reference shown in the table suggests that the sandwich variance estimator in  (\ref{eq:sandwich}) provides reliable estimation for the variance of the MCCL estimator. This settles the second issue.
\begin{table}[h]
	\caption{\label{tab:tab2}Averages of standard deviation estimates, $\widehat{\text{s.d.}}$, and empirical standard deviation, s.d., across $1000$ Monte Carlo replicates from (M1) with $\sigma^2_u=1.2$ and $n=200$. Numbers in parentheses are Monte Carlo standard errors associated with the Monte Carlo means.}
	\centering
	\begin{tabular}[t]{l*{8}{c}}
		\toprule
		& \multicolumn{2}{c}{$\beta_0$} & \multicolumn{2}{c}{$\beta_1$} & \multicolumn{2}{c}{$\beta_2$} & \multicolumn{2}{c}{$\log m$} \\
		\cmidrule(l{3pt}r{3pt}){2-3} \cmidrule(l{3pt}r{3pt}){4-5} \cmidrule(l{3pt}r{3pt}){6-7} \cmidrule(l{3pt}r{3pt}){8-9}
		& $\widehat{\text{s.d.}}$ & s.d. & $\widehat{\text { s.d. }}$ & s.d. & $\widehat{\text{s.d.}}$ & s.d. & $\widehat{\text{s.d.}}$ & s.d.\\
		\midrule
		$\mbox{MCCL}$ & 0.19 (0.03) & 0.19 & 0.13 (0.03) & 0.13 & 0.32 (0.06) & 0.32 & 0.15 (0.02) & 0.16\\
		Naive  & 0.16 (0.02) & 0.16 & 0.09 (0.01) & 0.09 & 0.26 (0.03) & 0.26 & 0.15 (0.01) & 0.16\\
		\bottomrule
	\end{tabular}
\end{table}

The third issue concerns the normality assumption on measurement error in the development of the Monte Carlo corrected score method. To assess the robustness of the MCCL estimator to this normality assumption, we revise (M1) by letting $U_{j,k}\sim \mbox{Laplace}(0, 0.5^{1/2})$ instead, \revise{for $k=1, 2, 3$}, and set $n=200$. Table \ref{tab:tab3} provides summary statistics of parameter estimates as those shown in Table~\ref{tab:tab1} \revise{(with $B=100$)} under this revised setting. \revise{In addition to estimates parallel to those considered in Table~\ref{tab:tab1}, we also include summary statistics for MCCL estimates obtained without using replicate measures of $X_1$. That is, we  keep $W_{j,1}$ in $\widetilde W_j =\{W_{j,k}\}_{k=1}^3$ as the only available error-contaminated measure of $X_{1,j}$, for $j=1, \ldots, 200$, when constructing the corrected log likelihood function. In Section~\ref{application-to-adni-data}, we describe a modified version of the correct log likelihood in (\ref{eq:correctedloglkh}) that does not require replicate measures but depends on the measurement error variance (see (\ref{eq:cs2002})). This creates a scenario where the violation of normality assumption associated with the measurement error in $W_{j,1}$ is more severe than when $\overline{W}_j=\sum_{k=1}^3 W_{j,k}/3$ is used as a surrogate of $X_{1,j}$.} As one can see from Table \ref{tab:tab3}, despite the \revise{(severity in)} violation of the normality assumption on $U$, the MCCL estimates remain close to the truth and significantly outperform the naive estimates. This robustness feature of the Monte Carlo corrected score method is also noted and explained in \citet{Steven2002}. 
\begin{table}[h]
	\caption{\label{tab:tab3}Medians of MCCL estimates and medians of naive estimates across $1000$
		Monte Carlo replicates generated according to (M1) with $U_{j,k}\sim \mbox{Laplace}(0, 0.5^{1/2})$ and $n=200$. The number in parentheses following each median is the interquartile range of the 1000 realizations of an estimator. \revise{$\text{MCCL}_{1}$ and $\text{MCCL}_{2}$ refer to MCCL estimates when replicate measures are present and absent, respectively.}}
	\centering
	\begin{tabular}[t]{l*{5}{c}}
		\toprule
		& $\beta_0$ & $\beta_1$ & $\beta_2$ & $\log m$\\
		\midrule
		$\text{MCCL}_{1}$ & 0.25 (0.26) & 0.25 (0.17) & 0.26 (0.41) & 1.12 (0.19)\\
		$\text{MCCL}_{2}$ & 0.25 (0.26) & 0.25 (0.17) & 0.26 (0.41) & 1.12 (0.19)\\
		Naive  & 0.18 (0.22) & 0.19 (0.13) & 0.39 (0.36) & 1.10 (0.19)\\
		\bottomrule
	\end{tabular}
\end{table}

\subsection{Performance of the model diagnostic method}\label{sec:modeldiag}
\begin{figure}[!h]
	{\centering \includegraphics{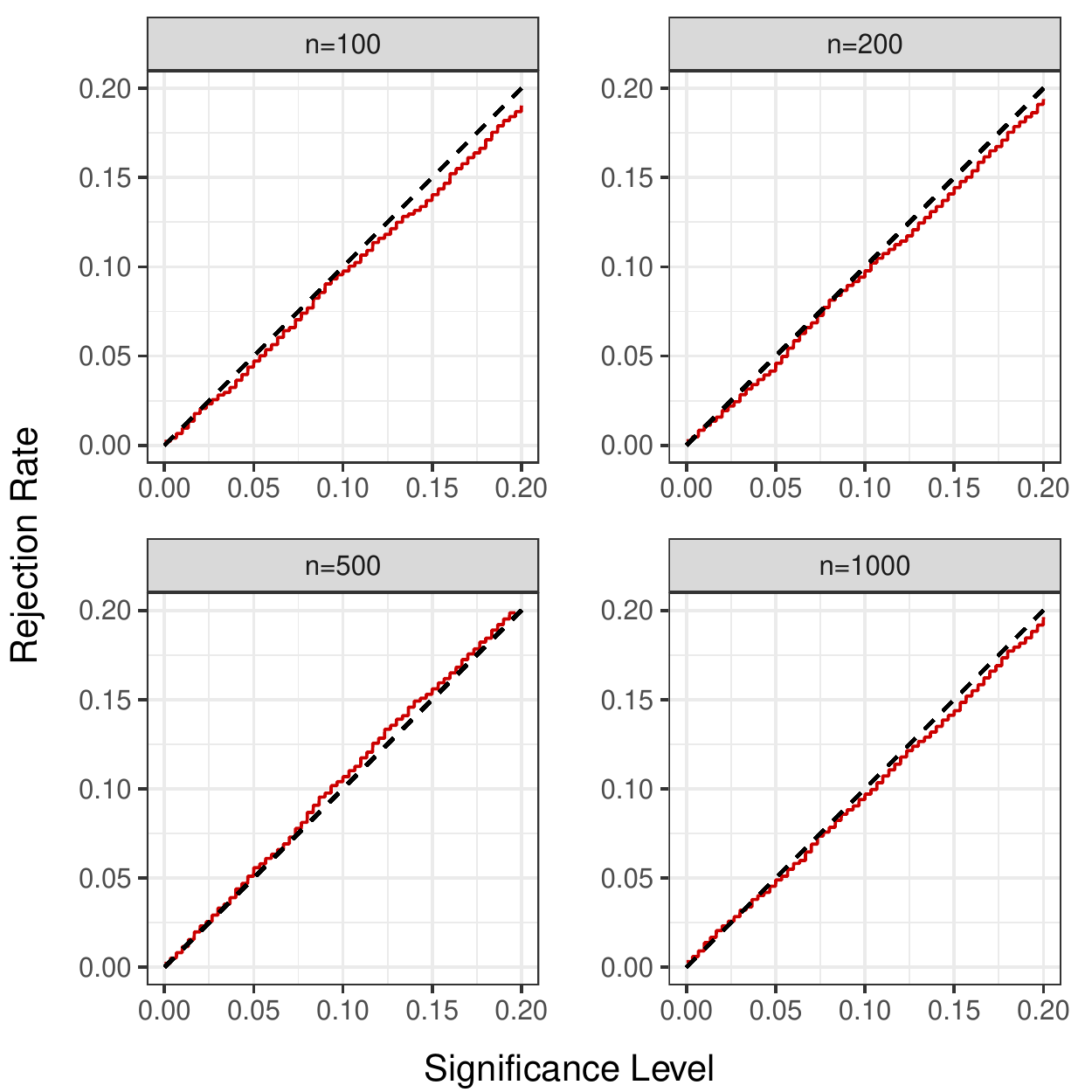} 
	}
	\caption{\label{fig:diagnosis}Rejection rates associated with the score-based diagnostic test across 5000 Monte Carlo replicates from (M1) versus the nominal level of the test. Black dashed lines are the $45^\circ$ reference lines.}
\end{figure}

Using 5000 Monte Carlo replicates from (M1) with $\sigma_u^2=1.2$ at each sample size level in $n=100$, 200, 500, 1000, we implement the bootstrap algorithm related in Section~\ref{sec:diagnos} with $M=300$ bootstrap samples to obtain estimated $p$-values associated with the test statistic $\tilde Q(\hat \bOmega; \mathcal{D}^*)$. We then record the proportion of replicates, across 5000 replicates, that lead to rejection of the null hypothesis of no model misspecification at various nominal levels. This rejection rate can be viewed as an empirical size of the test at a pre-specified significance level. Figure~\ref{fig:diagnosis} depicts this rejection rate versus the significance level, from which one can see that the size of the test is well controlled by the bootstrap procedure over a wide range of nominal levels. 

Table~\ref{tab:violation} presents rejection rates of the model diagnostic method in the presence of different forms of model misspecification that occur when fitting data generated according to (M2)--(M4) while assuming a beta modal regression model specified in (M1). As one can see in Table~\ref{tab:violation}, the proposed score-based test has moderate power to detect a misspecified form of the linear predictor, with the power steadily increasing as $n$ increases, and is especially powerful in detecting violation of the distributional assumption on $Y$ given covariates; but the test is less sensitive to link misspecification. Low power of most goodness-of-fit tests to detect link misspecification have been reported in the context of generalized linear models  \citep[e.g.,][]{hosmer1997comparison}. Given these reported findings in the literature, the low power observed under design (M3)  may not be surprising, especially with the high similarity of the logit link in the assumed model with the probit link in the true model in (M3). 
\begin{table}[!h]
	\caption{\label{tab:violation}Rejection rates of the score-based diagnostic test resulting from 300 Monte Carlo replicates in the presence of four types of model misspecification in (M2)--(M4)}
	\centering
	\begin{tabular}[t]{*{5}{c}}
		\toprule
		Model & $n=200$ & $n=300$ & $n=400$ & $n=500$\\
		\midrule
		(M2) & 0.283 & 0.407 & 0.550 & 0.580\\
		(M3) & 0.053 & 0.120 & 0.090 & 0.113\\
		(M4) & 1.000 & 0.997 & 0.997 & 1.000\\
		\bottomrule
	\end{tabular}
\end{table}

\revise{When the assumed beta modal regression model is rejected by the proposed diagnostic test, one may consider a more flexible unimodal distribution for the response conditioning on true covariates, such as the unimodal distributions formulated in \citet{fernandez1998bayesian}, \citet{quintana2009flexible}, \citet{rubio2015bayesian}, and \citet{liu2022flexible}. A different assumed primary regression model leads to a different log likelihood function $\ell(\bOmega; Y_j, \bX_j)$ in (\ref{eq:correctedloglkh}), and our proposed strategy of correcting a naive log likelihood function to account for measurement error remains applicable for any parametric regression models. }

\section{Real-life data application}\label{sec:real}
In this section, we analyze data arising from two different applications where a covariate of interest cannot be observed directly. Besides dealing with scientific questions in relevant fields, these applications provide opportunities for us to address some practical issues one faces when implementing the proposed estimation method and diagnostic method not discussed in the simulation study. 

\hypertarget{application-to-dietary-data}{%
	\subsection{Application to dietary
		data}\label{application-to-dietary-data}}
Food Frequency Questionnaire (FFQ) is a convenient and inexpensive dietary assessment instrument in epidemiologic studies. To study the association between an individual's FFQ intake  and his/her long-term usual intake as the univariate covariate \(X\), we analyze a dietary data set from Women's Interview Survey of Health \citep{Carroll1997}.
The data set contains 271 females' FFQ intake records, measured as the percentage calories from fat, and six \(24\)-hour food recalls, \(W_{j,k}\), for \(j = 1,\ldots, 271\) and \(k = 1,\cdots, 6\). Because the $j$-th subject's long-term usual intake $X_j$ cannot be measured directly, a generally accepted practice in epidemiology is to use  \(\overline{W}_j=\sum_{k=1}^6 W_{j,k}/6\) as a surrogate of \(X_j\), for $j=1, \ldots, 271$. According to the preliminary analysis in existing literature, the distribution of the FFQ intake appears to be right-skewed and potentially heavy-tailed, which motivates the consideration of a modal regression model in place of a mean regression model. Here, we assume a beta modal regression model given in (\ref{eq:yx}) with $\theta(X)=1/\{1+\exp(-\beta_0-\beta_1 X)\}$ for the response data $\{Y_j\}_{j=1}^{271}$, where $Y_j$ is the $j$-th subject's FFQ intake in kilocalorie divided by 8000, a biologically plausible upper bound of daily energy intakes for a general population. 

We obtain the MCCL estimate of $\bOmega=(\beta_0, \beta_1, \log m)^\T$ according to (\ref{eq:mestimatorfast}), and also carry out regression analysis that ignores measurement error to obtain a naive maximum likelihood estimate of $\bOmega$. \revise{Moreover, we implemented the simulation-extrapolation method \citep[SIMEX,][Chapter 5]{Carroll2006} applied to the assumed beta modal regression model. In this particular application, SIMEX amounts to repeatedly estimating $\bOmega$, without accounting for measurement error, using data $\mathcal{D}_b^*(\zeta)=\{(Y_j, W_{j,b}(\zeta))\}_{j=1}^ n$, for $b=1, \ldots, B$, where $W_{j,b}(\zeta)=\overline W_j+\sqrt{\zeta} \sigma_u Z_{j,b}$, in which $\{Z_{j,b}, j=1, \ldots, n\}_{b=1}^{B}$ are independent standard normal errors, $\sigma_u$ is the standard deviation of measurement error associated with the surrogate measure $\overline W_j$, and $\zeta$ is a user-specified positive constant. Denote by $\hat\bOmega_b(\zeta)$ the (naive) estimator of $\bOmega$ based on data $\mathcal{D}^*_b(\zeta)$, then $\hat\bOmega(\zeta)=\sum_{b=1}^B \hat\bOmega_b(\zeta)/B$ is a naive estimator based on data resulting from further contaminating the original error-prone data $\mathcal{D}^*=\{(Y_j, \overline W_j)\}_{j=1}^n$, with the amount of additional contamination controlled by $\zeta$. Collecting a sequence of $\hat \bOmega(\zeta)$ as one varies $\zeta$ realizes the simulation step of SIMEX. In this data application, we set $B=300$ and let $\zeta$ vary from 0.125 to 1 in increments of 0.125. The extrapolation step of SIMEX entails extrapolating the sequence of estimates in $\{\hat \bOmega(\zeta), \, \zeta=0.125, 0.25, \ldots, 1\}$ to $\hat \bOmega(-1)$, leading to the so-called SIMEX estimator. A heuristic motivation of extrapolating towards $\zeta=-1$ can be revealed by noting that $\text{Var}(W_{j,b}(\zeta)|X_j)=\text{Var}(\overline W_j|X_j)+\zeta \sigma_u^2$, where $\sigma_u^2=\text{Var}(\overline W_j|X_j)$. Setting $\zeta=-1$ in the preceding variance expression gives $\text{Var}(W_{j,b}(-1)|X_j)=0$, as if $W_{j,b}(-1)$ contained no measurement error, and hence extrapolating $\{\hat \bOmega(\zeta), \text{ for $\zeta>0$}\}$ to obtain $\hat \bOmega(-1)$ is an attempt to ``recover" an estimator of $\bOmega$ had there been no covariate measurement error. \citet{Shi2021} applied SIMEX to a kernel-based modal regression model with error-prone covariates.}  

\begin{figure}
	{\centering \includegraphics[width = 1\textwidth]{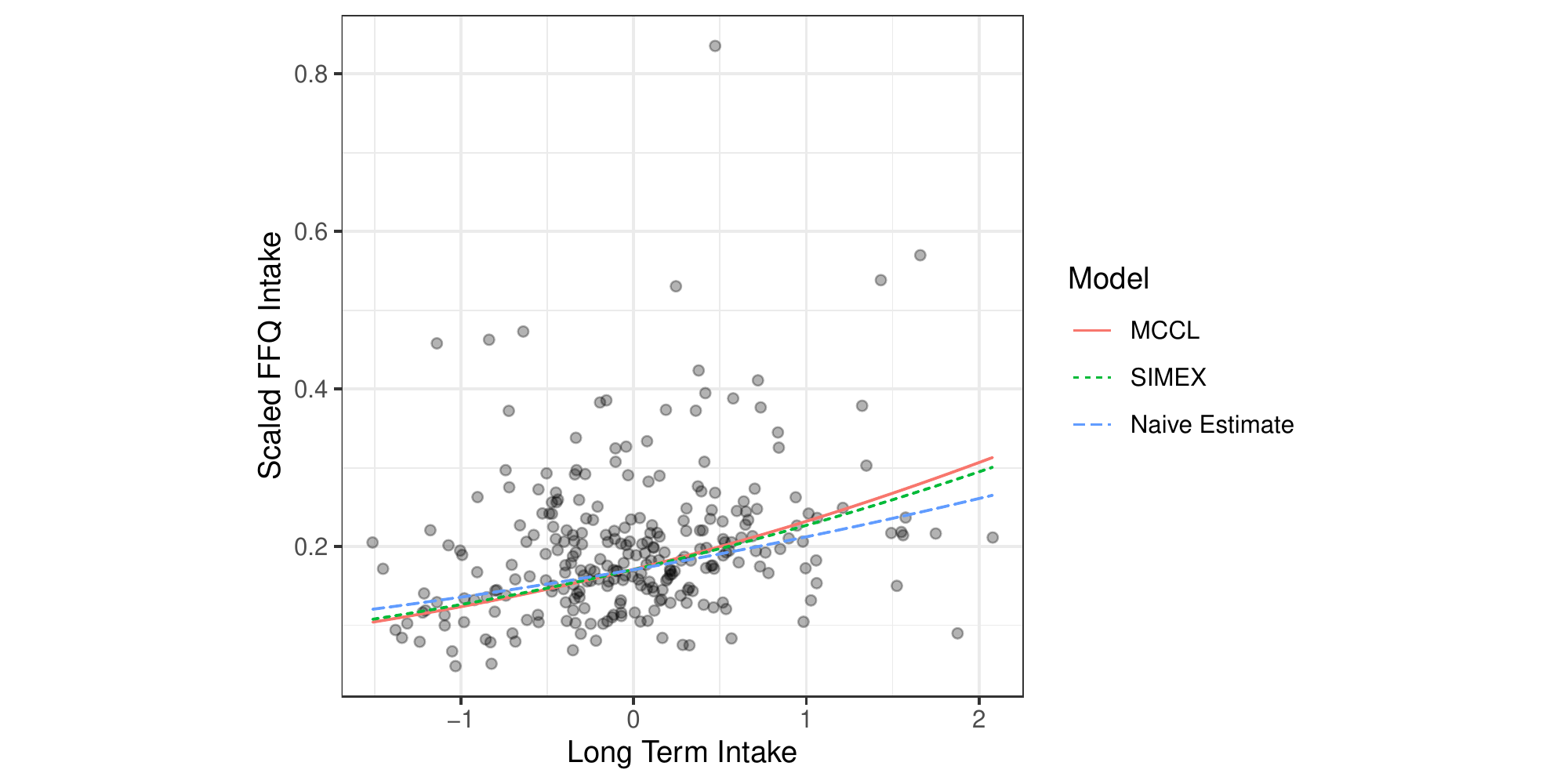} 
	}
	\caption{Estimated conditional mode functions for  the dietary data based on the MCCL estimate (red solid line) and the naive estimate (cyan dashed line), respectively. Observed covariate data $\{\overline{W}_j\}_{j=1}^{271}$ are treated as surrogates of long-term usual intakes in the scatter plot of the observed data (solid dots).}\label{fig:FFQ}
\end{figure}

\revise{Three estimates of $\bOmega$, the MCCL estimate, SIMEX estimate, and naive estimate,} are given in Table~\ref{tab:FFQ}. The covariate effect associated with the long-term intake suggested by the naive estimate is substantially weaker than that indicated by the MCCL estimate \revise{and SIMEX estimate}, implying potentially significant attenuation on the covariate effect due to measurement error in the former, whereas the latter \revise{two correct} for this attenuation. Figure \ref{fig:FFQ} depicts the estimated regression functions $\hat \theta(x)$ resulting from these \revise{three} methods, imposed on the scaled response data versus the surrogate covariate data. This pictorial contrast between the \revise{three} estimated regression functions shows that the proposed method \revise{and SIMEX are} able to capture the underlying positive non-linear covariate effect that is partially concealed or weakened by the naive method. \revise{Although SIMEX produces similar inference results as those from our method, the simulation step relies on the error variance $\sigma^2_u$ when generating $W_{j,b}(\zeta)$'s, which we estimate in this example based on replicate measures; and the extrapolation step depends on the choice of an extrapolant, a choice that usually lacks data evidence to support in most applications. Here, we use a quadratic extrapolant to obtain the SIMEX estimate. Besides being more computationally burdensome compared to the MCCL method (due to repeatedly estimating $\bOmega$ based further contaminated data), variance estimation for SIMEX estimators is also less straightforward than that for our estimator \citep{carroll1996asymptotics}. We resort to nonparametric bootstrap, with 1000 bootstrap samples, in this example to obtain the estimated standard errors associated with SIMEX estimates shown in Table~\ref{tab:FFQ}.} Finally, applying the proposed diagnostic method to this data set with $M=300$ bootstrap samples yields an estimated $p$-value of 0.097. We thus conclude lack of sufficient data evidence (at significance level 0.05) to indicate the assumed beta modal regression model inadequate for this application. 
\begin{table}[h]
	\caption{\label{tab:FFQ}Estimates of parameters in the beta modal regression model applied to the dietary data, along with the corresponding estimated standard errors in parentheses}
	\centering
	\begin{tabular}[t]{*{4}{c}}
		\toprule
		Method & $\beta_0$ & $\beta_1$ & $\log m$ \\
		\hline
		MCCL & $-1.578$ (0.033) & 0.381 (0.099) & 3.015 (0.196) \\
		\revise{SIMEX} & $-1.580$ (0.034) & 0.354 (0.087) & 3.008 (0.195) \\
		Naive & $-1.581$ (0.041) & 0.270 (0.058) & 2.979 (0.094) \\
		\bottomrule
	\end{tabular}
\end{table}

\hypertarget{application-to-adni-data}{%
	\subsection{Application to Alzheimer's disease  data}\label{application-to-adni-data}}
Medical researchers have long recognized that cerebral atrophy is associated with dementia, and extensive research have been conducted to understand the association between volumetric changes of different brain regions with the severity of dementia. Abundant data collected from this line of research are available in the Alzheimer's Disease Neuroimaging Initiative (ADNI) database (\url{http://adni.loni.usc.edu/}). \citet{zhouhuang2020} analyzed a data set relating to 245 individuals diagnosed with mild cognitive impairment from this database. The goal is to study roles that an individual's volumetric measure of entorhinal cortex (ERC) and that of hippocampus (HPC) play in predicting one's risk of developing Alzheimer's disease. An individual's test score from the Alzheimer’s disease assessment scale, known as  ADAS-11, at month 12 since entering the ADNI cohort is used to assess one's severity of cognitive impairment. Covariates of interest are the volumetric change in ERC (ERC.change) and that in HPC (HPC.change) at month 12 compared to the baseline measures collected at month 6. Assuming these volumetric measures are observed precisely, \citet{zhouhuang2020} fitted the data  to the beta modal regression model for the response $Y$ defined as an individual's ADAS-11 score divided by a perfect score of 70, with the log-log link in the mode function, $\theta(\bX)=\exp\{-\exp(-\beta_0-\beta_1 \times \text{ERC.change}-\beta_2\times \text{HPC.change})\}$, and showed that it provides a better fit for the data compared to the  beta mean regression model proposed by \citet{Silvia2004}. 

In reality, measuring ERC volume is challenging because of lateral border discrimination from the perirhinal cortex \citep{price2010}, and the accuracy of HPC measurements is also in question \citep{Maclaren2014}. It is thus more sensible to view the observed volumetric change of ERC or that of HPC as a noisy surrogate of the actual amount of change. Despite of which covariate is viewed as error-prone, the current data present some challenges due to the lack of replicate measures for an individual's true covariate value, and thus the estimation methods proposed in Section~\ref{sec:estimation} are not applicable. For example, in (\ref{eq:correctedloglkh}), the term multiplying the imaginary unit $i$ is equal to zero now with the number of replicates $n_j=1$, making the ``corrected" log-likelihood the same as the naive log-likelihood. A quick fix to the problem is to invoke a similar strategy of correcting naive scores to account for measurement error as discussed in \citet{Steven2002}. Following this strategy, a corrected log-likelihood evaluated at the $j$-th data point to use in place of (\ref{eq:correctedloglkh}) is 
\begin{equation}
	\tilde \ell(\bOmega; Y_j, \bW_j, \tilde \bZ_j)=\frac{1}{B}\sum_{b=1}^B \text{Re}\{\ell(\bOmega; Y_j, \bW_j+i\bSigma_u^{1/2} \bZ_{j,b})\},
	\label{eq:cs2002}
\end{equation}
where $\tilde \bZ_j=\{\bZ_{j,b}\}_{b=1}^B$, for $j=1, \ldots, n$, and $\{\bZ_{j,b}, b=1, \ldots, B\}_{j=1}^n$ are independent $p$-dimensional normal random vectors with mean zero and variance-covariance as an identity matrix, which accommodates multiple error-prone covariates in $\bX$ by letting $\bW_j$ be a $p$-dimensional multivariate surrogate of $\bX_j$, contaminated by a multivariate normal measurement error $\bU_j$ with variance-covariance matrix  $\bSigma_u$. By setting all entries in $\bSigma_u$ at zero except for the first diagonal entry gives rise to the case considered in the majority of this article with only $X_1$ prone to error. Certainly, not having replicate measures still creates an obstacle to implementing this strategy due to its dependence on $\bSigma_u$ that cannot be estimated without replicate measures of a true multivariate covariate value or other external validation data. A well-accepted practice among statisticians in similar situations is to carry out sensitivity analysis where one analyzes the data under different assumptions for the parameter, such as $\bSigma_u$ in our case, that one lacks data information to infer. If one obtains drastically different inference results when assuming different values for $\bSigma_u$, including a matrix of zeros corresponding to naive estimation that ignores measurement error, then one may recommend to exercise caution when interpreting results from an inference procedure that assumes error-free covariates. 

For illustration purposes, we assume in the sensitivity analysis four values for $\bSigma_u$ listed in Table \ref{tab:ADNI}, where inference results for model parameters under each assumed $\bSigma_u$ are provided. According to Table~\ref{tab:ADNI}, all four rounds of regression analyses lead to the conclusion that the volumetric change of ERC is an influential predictor for the severity of cognitive impairment, even though the magnitude of the estimated covariate effect is sensitive to the assumed error variance associated this covariate. In particular, when assuming imprecise measurements for ERC.change, the revised MCCL method that employs the corrected log-likelihood in (\ref{eq:cs2002}) with $B=100000$ produces results indicating a much stronger association than the naive analysis. By comparison, the magnitude of the estimate for the HPC.change effect is less sensitive to the assumed $\bSigma_u$, but its statistical significance is noticeably affected by it. For example, one would conclude a moderately significant covariate effect of HPC.change based on the naive analysis assuming error-free covariates, but claim a highly significant, or moderately significant, or nonsignificant HPC.change effect depending on which covariate(s) one assumes to be error-prone and the severity of error contamination. This phenomenon is a reminiscence of an observation made in Figure \ref{fig:fig1}, and may suggest that ERC.change and HPC.change are correlated. In fact,  measurements of ERC and HPC via magnetic resonance imaging are known to be highly correlated with observed clinical alterations in patients suffering mild cognitive impairment or at dementia phases of Alzheimer's disease \citep{Desikan2010,  Jack2013, Varon2014}. 
\begin{table}[h]
	\caption{\label{tab:ADNI}Sensitivity analysis using the ADNI data for the beta modal regression with the log-log link. Numbers in parentheses are estimated standard errors. Numbers in square brackets are $p$-values associated with covariate effects.}
	\centering
	\begin{tabular}{ccccc}
		\toprule
		$\bSigma_u$ & $\beta_0$ & $\beta_1$ (ERC.change) & $\beta_2$ (HPC.change) & $\log m$ \\
		\midrule
		\multirow{2}{*}{
			$\begin{bmatrix}
				0 & 0 \\
				0 & 0
			\end{bmatrix}$} & $-0.69$ (0.03) & $-0.12$ (0.05) & $-0.22$ (0.11) & 2.78 (0.15) \\
		&                & [0.007]        & [0.054]         & \\
		\multirow{2}{*}{
			$\begin{bmatrix}
				0.16 & 0 \\
				0 & 0 
			\end{bmatrix}$} & $-0.88$ (0.03) & $-2.44$ (0.00) & $0.39$ (0.46) & 3.45 (0.04) \\
		&                & [0.000]        & [0.386]        & \\
		\multirow{2}{*}{
			$\begin{bmatrix}
				0 & 0 \\
				0 & 0.0225 
			\end{bmatrix}$} & $-0.71$ (0.03) & $-0.11$ (0.05) & $-0.47$ (0.27) & 2.80 (0.16) \\
		&               & [0.014]        & [0.084]        & \\
		\multirow{2}{*}{
			$\begin{bmatrix}
				0.16 & 0 \\
				0 & 0.0225 
			\end{bmatrix}$} & $-0.81$ (0.03) & $-2.42$ (0.00) & $-0.85$ (0.00) & 3.85 (0.02) \\
		&                & [0.000]        & [0.000]        & \\
		\bottomrule
	\end{tabular}
\end{table}

In conclusion, results from the sensitivity analysis suggest that volumetric measures of different brain regions are likely to be subject to measurement error, and statistical analyses under the assumption of precisely measured covariates should be interpreted with caution. If replicate data are available for covariates of interest, the MCCL method can provide more reliable inference. Lastly, even though one can mimic (\ref{eq:cs2002}) to construct a corrected score in place of $\tilde \bS(\bOmega; Y_j, \widetilde{W}_j, \widetilde{T}_j, \bX_{-1,j})$ in (\ref{eq:Stilde}) and then formulate the test statistic $\tilde Q(\hat \bOmega; \mathcal{D}^*)$ for model diagnostics, the dependence of the revised score on the unknown $\bSigma_u$ remains an obstacle that hinders one from using the bootstrap procedure outlined in Section~\ref{sec:diagnos} to assess statistical significance of the revised test statistic. Alternative diagnostic methods that do not rely on parametric bootstrap or corrected score \citep[e.g.][]{huang2006latent} can be used to detect inadequate assumptions imposed on the primary regression model. 

\section{Discussion}\label{sec:discussion}
We propose an inference procedure based on the idea of corrected score that falls in the framework of $M$-estimation for modal regression with an error-prone covariate. Even though in this article we focus on the  beta modal regression model as the primary regression model, the proposed MCCL method is applicable in other parametric modal regression models, such as the gamma modal regression models for non-negative responses proposed by \citet{aristodemou2014new} and \citet{bourguignon2020parametric}\revise{, 
	and the flexible Gumbel regression model  recently proposed by \citet{liu2022flexible} for responses ranging over the entire real line. In fact, provided that a parametric modal regression model can provide reliable inference for the global mode in the absence of covariate measurement error (even when $Y$ follows a multimodal distribution given $\bX$), such as the flexible unimodal regression models considered in \citet{liu2022bayesian}, the proposed MCCL method applied to error-prone data is expected to improve over the counterpart naive method that ignores measurement error.} A Python package for implementing the proposed methods for beta modal regression with errors-in-covariate is available at  
\url{https://pypi.org/project/pybetareg/}. All computer programs used in this paper are available at \url{https://github.com/rh8liuqy/Modal_regression_with_measurement_error}.

To accommodate situations without replicate measures of the true covariate or settings with multiple error-prone covariates, the MCCL method can be easily revised as demonstrated in Section~\ref{application-to-adni-data}, although one needs to specify the variance (or the variance-covariance matrix) of the (vector-valued) measurement error if one lacks replicate data or external validation data to estimate it. 

Focusing on the current beta modal regression models, some extensions are worthy of further investigation, such as a zero-inflated beta modal regression model to fit disease prevalence data especially suitable for rare diseases, \revise{and a four-parameter beta modal regression model as considered in \citet{zhouhuang2020} for a bounded response with unknown support}. Another follow-up research direction is variable selection based on a parametric modal regression model with or without measurement error contamination in covariates. 

\vspace*{1pc}

\noindent {\bf{Conflict of Interest}}

\noindent {\it{The authors have declared no conflict of interest. }}

\bibliographystyle{apalike}
\bibliography{mybibfile}
	
\end{document}